      [1999/12/01 v1.4c Il Nuovo Cimento]
\documentclass{cimento}
\usepackage{amsmath}
\usepackage{epsfig}
\usepackage{xspace}
\usepackage{cite}
\usepackage{amssymb}
\usepackage{wrapfig}
\usepackage{calc}
\usepackage{mathptmx}

\usepackage{graphicx}  

\title{Di-hadron correlations at RHIC}
\author{C.~Nattrass\from{ins:UT}\ETC}
\instlist{\inst{ins:UT}  University of Tennessee at Knoxville\-Knoxville, Tennessee, USA}
\PACSes{{25.75.-q,21.65.Qr,24.85.+p,25.75.Bh}}

\newcommand{\sNNtwohundred}{$\sqrt{s_{NN}}$ = 200 GeV\xspace}
\newcommand{\sNNsixtytwo}{$\sqrt{s_{NN}}$ = 62 GeV\xspace}

\newcommand{\stdassoc}{$1.5$ GeV/c $<$ $p_T^{associated}$ $<$ $p_T^{trigger}$\xspace}
\newcommand{\stdtrig}{$3.0$ $<$ $p_T^{trigger}$ $<$ $6.0$ GeV/c\xspace}
\newcommand{\pttrig}{$p_T^{trigger}$\xspace}
\newcommand{\ptassoc}{$p_T^{associated}$\xspace}
\newcommand{\trigrange}[2]{#1 $< p_T^{trigger} < $ #2 GeV/c\xspace}

\newcommand{\assocrangevar}[1]{#1 $< p_T^{associated} < p_T^{trigger}$ GeV/c\xspace}

\newcommand{\npart}{$N_{part}$\xspace}

\newcommand{\pT}{$p_T$\xspace}

\newcommand{\RAA}{$R_{AA}$\xspace}

\newcommand{\highpT}{high-$p_T$\xspace}

\newcommand{\ridge}{ridge\xspace}

\newcommand{\pp}{$p+p$\xspace}

\newcommand{\Cu}{$Cu+Cu$\xspace}
\newcommand{\Au}{$Au+Au$\xspace}
\newcommand{\Pb}{$Pb+Pb$\xspace}
\newcommand{\dAu}{$d+Au$\xspace}
\newcommand{\AplusA}{$A+A$\xspace}

\newcommand{\GeV}{GeV/c\xspace}
\newcommand{\dphi}{$\Delta\phi$\xspace}
\newcommand{\deta}{$\Delta\eta$\xspace}
\newcommand{\vtwo}{$v_2$\xspace}

\newcommand{\ns}{near-side\xspace}
\newcommand{\as}{away-side\xspace}
\newcommand{\baryontomeson}{$\frac{\Lambda+\bar{\Lambda}}{2K^0_S}$\xspace}
\newcommand{\ptopi}{$\frac{p+\bar{p}}{\pi^++\pi^-}$\xspace}

\newcommand{\Fref}[1]{Figure~\ref{#1}}

\newcommand{\Cref}[1]{Chapter~\ref{#1}}
\newcommand{\Eref}[1]{Equation \ref{#1}}

\newcommand{\jlc}{jet-like correlation\xspace}

\newcommand{\jly}{jet-like yield\xspace}

\newcommand{\ry}{\ridge yield\xspace}

\newcommand{\py}{PYTHIA\xspace}

\newcommand{\degree}{\ensuremath{^\circ}\xspace}

\newcommand{\lt}{$<$\xspace}

\def\Dphi{\mbox{$\Delta\phi$}}

\begin{document}

\maketitle

\begin{abstract}
Di-hadron correlations have been used to study jets at RHIC and have yielded rich insight into the properties of the medium.  Studies show that the near-side peak of \highpT triggered correlations can be decomposed into two parts, a jet-like correlation and the ridge.  The jet-like correlation is narrow in both azimuth and pseudorapidity and has properties consistent with vacuum fragmentation, while the ridge is narrow in azimuth but broad in pseudorapidity and roughly independent of pseudorapidity.  The energy, system, and particle composition of the jet-like correlation and the ridge are discussed.  Data indicate that the jet-like correlation is dominantly produced by vacuum fragmentation.  Attempts have been made to explain the production of the ridge component as coming from recombination, momentum kicks, and QCD magnetic fields.  However, few models have attempted to quantitatively calculate the characteristics of the ridge.  The wealth of data should help distinguish models for the production mechanism of the ridge.  Implications for studies of the jet-like correlation and the ridge at the LHC are discussed.
\end{abstract}

Measurements of the suppression of \highpT hadrons in \AplusA relative to \pp at the Relativistic Heavy Ion Collider (RHIC) demonstrate that there is strong suppression of \highpT hadrons in the presence of a hot, dense medium \cite{Adare:2008qa,Abelev:2007ra}.  Di-hadron correlations have been used at RHIC as another way of measuring jet suppression.

\begin{figure}
  \centering\resizebox{14cm}{!}{\includegraphics{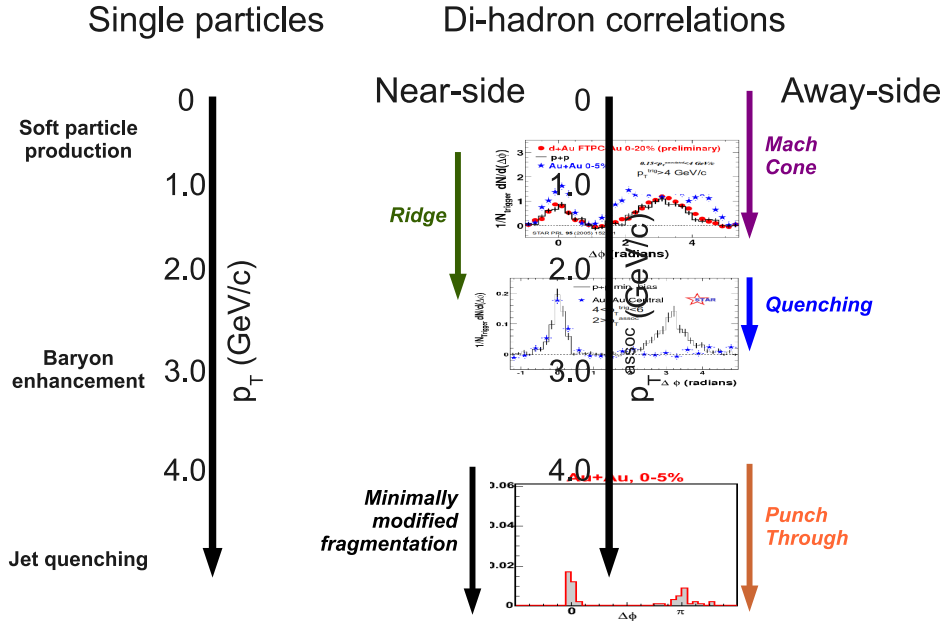}}\\

  \caption{Schematic diagram showing the different kinematic regions for single hadrons and for di-hadron correlations.  Figures are from \cite{STARLowPtAssoc2005,STARIconicJetQuench2003,STARPunchThrough}  Colour online.}\label{SchematicFigure}
\end{figure}

\begin{figure}
  \centering\resizebox{7cm}{!}{\rotatebox{-90}{\includegraphics{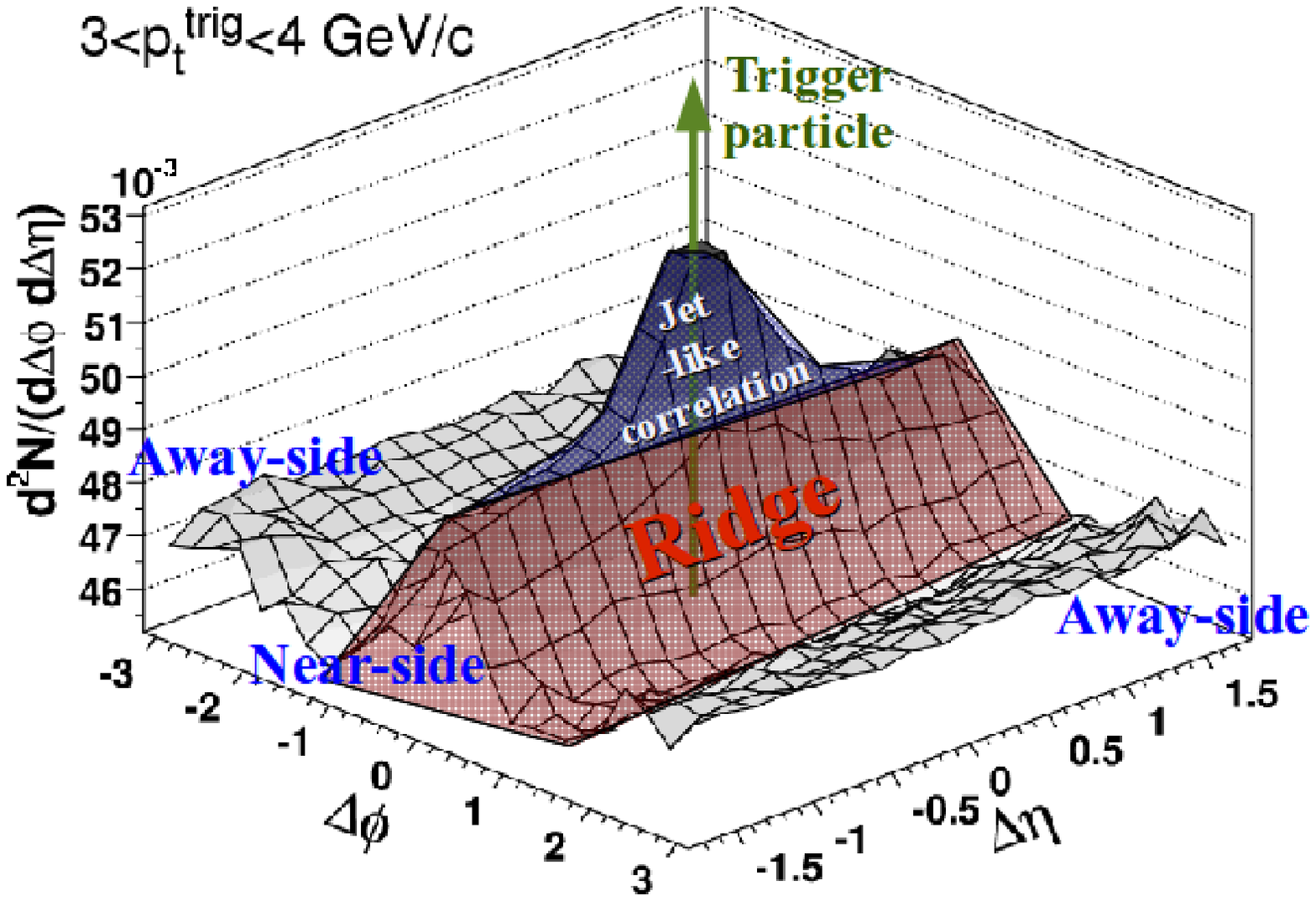}}}\\
  \caption{Data from 0-12\% central \Au collisions at \sNNtwohundred from \cite{STARRidgePaper} with \trigrange{3}{4} and \assocrangevar{2} with the \ridge, the \jlc, the \as and the location of the trigger particle labeled.  Colour online.}\label{STARRidge}
\end{figure}

In a standard di-hadron correlation analysis, a \highpT trigger particle is selected and the distribution of associated particles relative to this trigger particle is determined.  The correlation on the same side as the trigger particle is called the \ns and the correlation 180\degree from the trigger particle is called the \as.
Several features have been observed in di-hadron correlations, all of which are highly dependent on the kinematic region studied.  
The different kinematic regions are shown schematically in \Fref{SchematicFigure} with the different effects observed at RHIC for both inclusive particle ratios and di-hadron correlations.  
At intermediate \pT (2-6 GeV/c) baryon enhancement is observed in the inclusive particle ratios \cite{Abelev:2007ra,Adler:2003cb}.  
At higher momenta, baryon to meson ratios in \AplusA collisions are comparable to those observed in \pp collisions \cite{Abelev:2007ra} and the inclusive \RAA shows a strong suppression of \highpT hadrons \cite{Adare:2008qa,Abelev:2007ra}.  
On the \as at low \ptassoc, there is dip at roughly 180\degree in azimuth away from the trigger particle \cite{Aggarwal:2010rf,PHENIXHeadAndShoulders}.  This is often referred to as the Mach Cone after one of the models for the formation of this structure where a hard parton moving faster than the speed of sound in the medium creates a shock wave \cite{Renk:2005si}, although there are other models for this structure \cite{Vitev:2005yg,Polosa:2006hb,Chiu:2006pu,Dremin:2005an}.  For \ptassoc $\gtrapprox$ 2 \GeV, the \as is suppressed.  On the \ns for 1 \lt \ptassoc \lt 3 \GeV, there are two structures, shown in \Fref{STARRidge}.  The \jlc is narrow in both pseudorapidity (\deta) and azimuth (\dphi) and is present in \dAu, \Cu and \Au collisions.  The \ridge is a novel feature first observed in \Au collisions \cite{STARRidgePaper,Putschke:2007mi}.  For roughly 1 \lt \ptassoc \lt 3 \GeV, the \ridge is the dominant structure \cite{STARRidgePaper}, in the same kinematic region where we see baryon enhancement in the inclusive particle spectra.  The \ridge is also present at lower momenta \cite{PHOBOSRidge} and extends to at least \ptassoc $\approx$ 8 \GeV \cite{STARRidgePaper} and at least \deta=4 \cite{PHOBOSRidge}.  Above \ptassoc$\approx$ 3 \GeV the \jlc is the dominant structure on the \ns \cite{Oana:2009gg}.  On the \as, for higher \pttrig and \ptassoc the \as peak reappears \cite{STARPunchThrough}, in the same kinematic region where the inclusive particle ratios approach the values seen in \pp collisions.

%
%


%

\section{Experimental measurements}
The primary criterion used to determine trigger and associated particles is their momenta.  High-\pT triggered di-hadron correlations typically method neglect any correlations between high-\pT particles not caused by jets or anisotropic flow.  The momenta of the trigger and associated particles are restricted to high-\pT to increase the probability that the particles come from a jet and therefore decrease the combinatorial background.
The STAR collaboration presents correlations normalized per trigger particle \cite{STARLowPtAssoc2005,STARIconicJetQuench2003,STARIconicJetQuench2002}.  With this normalization, results from different systems (\pp, \dAu, \Cu, and \Au) would be identical if there were no modification of the jet.  The PHENIX collaboration uses both this normalization and a normalization where the amplitude of the correlation is interpreted as the probability for an associated particle to be correlated with the jet \cite{PHENIXPartIDCorr2007,PHENIXHeadAndShoulders,PHENIXCorrEnergyDependence}.  In addition to triggered di-hadron correlation measurements, some studies are untriggered and use the minimum \pT within the acceptance of the detector \cite{STARMiniJets2006,Daugherity:2008su,PHOBOSRidge}.  These analyses are typically normalized such that the amplitude scales with the probability that a random pair of particles with a separation (\dphi,\deta) are correlated, although the exact details of the normalization vary with the measurement.

\subsection{Background subtraction}\label{backgroundsubtraction}
There is a combinatorial background in this method from trigger and associated particles whose production is not correlated.  In \AplusA collisions, this background is modulated by anisotropic flow of particles in the medium, giving a background of the form \cite{PhysRevC.66.034904,Bielcikova:2003ku}
\begin{equation}
\label{eq:bkgd}
 b_{\Delta\phi}
\left(1+2 \langle v_{2}^{trig} v_{2}^{assoc} \rangle \cos2\Dphi+2 \langle v_{4}^{trig} v_{4}^{assoc} \rangle \cos4\Dphi+...\right) .
\end{equation}
\noindent
Odd order terms ($v_1, v_3...$) are assumed to cancel out because these terms are asymmetric with respect to the reaction plane.  Di-hadron correlation analyses are averaged over several events and the sign of $v_n$ is as likely to be positive as negative in each event for odd n so the average is roughly zero.  
For most analyses, only the \vtwo terms are considered and \vtwo is determined from independent analyses.  These independent analyses have large systematic errors because they may be affected by azimuthal anisotropies from sources other than hydrodynamical flow, such as jet production \cite{Voloshin:2008dg}.  The fact that flow leads to a background for studies of jets and jets lead to a background for studies of flow makes separating these effects complicated.  
There also may be event-by-event fluctuations in \vtwo, meaning that the average \vtwo in all events may not be the average \vtwo in the events used for the di-hadron correlation analysis.  STAR and PHENIX assumed that \vtwo is independent of $\eta$, which is valid in their acceptances \cite{Back:2004mh,Back:2004zg}, and PHOBOS takes the $\eta$ dependence of \vtwo into account for their background subtraction.  \vtwo is on the order of 0.1 and $0.75 v_2^2 < v_4 < 1.5 v_2^2$ \cite{Adare:2010ux} so $v_4$ terms are negligible in most analyses.

To determine the background for di-hadron correlation analyses, two additional assumptions are usually made.  The first assumption is that the raw signal comprises only two components, the combinatorial background as given by \Eref{eq:bkgd} and the signal.  This is generally called the Two-Component Model.  The raw signal (S) is assumed to come from particles from a fragmenting hard parton (J) and particles coming from coming from the combinatorial background (B):
\begin{equation}
\label{eq:schem}
S = J_1 J_2 + J_1 B_2 + J_2 B_1 + B_1 B_2.
\end{equation}
\noindent
In \AplusA the signal is much smaller than the background.  The cross terms are typically neglected.  If jet production is not correlated with the reaction plane, the cross terms $J_1 B_2 + J_2 B_1$ would add a constant background.  This corresponds to having a lower effective $\langle v_{2}^{trig} v_{2}^{assoc}\rangle$ \cite{Ajitanand:2005jj}.  Jet quenching could also lead to an azimuthal anisotropy of \highpT hadrons that would cause the cross terms to have the same form as \Eref{eq:bkgd}, but the azimuthal anisotropy would have a different magnitude and a different physical origin than the \vtwo from anisotropic flow.  The $B_1 B_2$ term is described by \Eref{eq:bkgd}.  These assumptions alone are not sufficient to determine the background because the level of the background needs to be fixed.

The most common assumption used to determine the level of the background is that there is a region in azimuth (\dphi$\approx$1) where there is no contribution from the signal.  A background of the form in \Eref{eq:bkgd} is fixed in this region, assuming that there are no other relevant terms.  This method is called the Zero-Yield-At-1 (ZYA1)\cite{STARLowPtAssoc2005} or Zero-Yield-At-Minimum (ZYAM) \cite{Ajitanand:2005jj} method.  Alternative methods have been proposed \cite{Sickles:2009ka}, however, these still require an assumption about the amount of combinatorial background in the raw signal.

Most analyses therefore have several inherent assumptions:
(1) The only combinatorial background is from particles correlated with the reaction plane due to hydrodynamical flow in the medium.
(2) The \vtwo term of the azimuthal anisotropy is the only relevant contribution from flow.
(3) Jet fragmentation is not correlated with the reaction plane.
(4) The minimum in the di-hadron correlation has no contributions from the signal.
In addition, studies generally consider the near- and \as separately.  These assumptions are generally a reasonable approximation when the background is small and the near- and \as peaks are well separated, corresponding to higher \pttrig and \ptassoc, but they are more ambigious in the intermediate \pT (2-4 \GeV) range.

\begin{figure}
  \centering\resizebox{14cm}{!}{
  \includegraphics{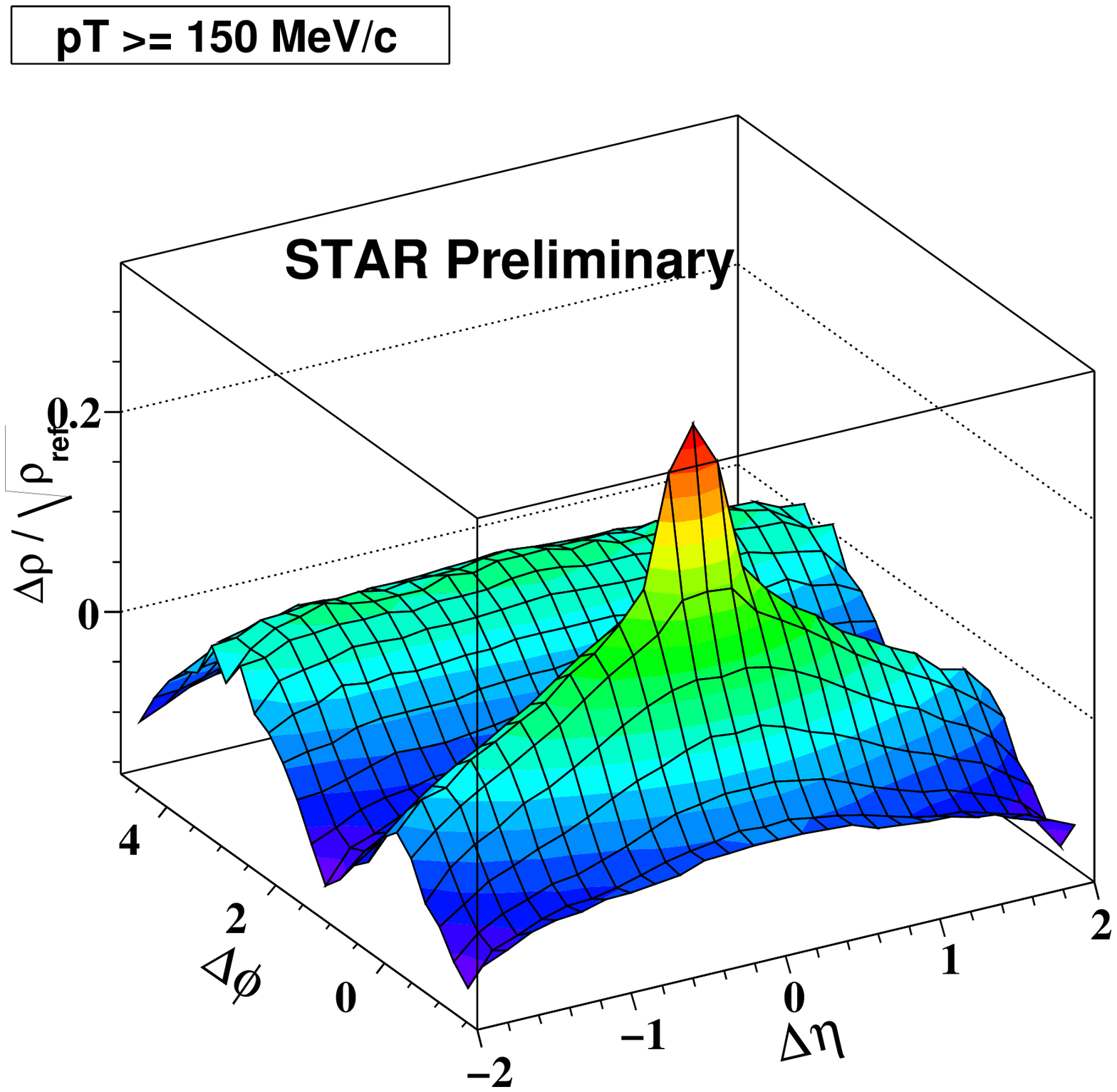}
  \includegraphics{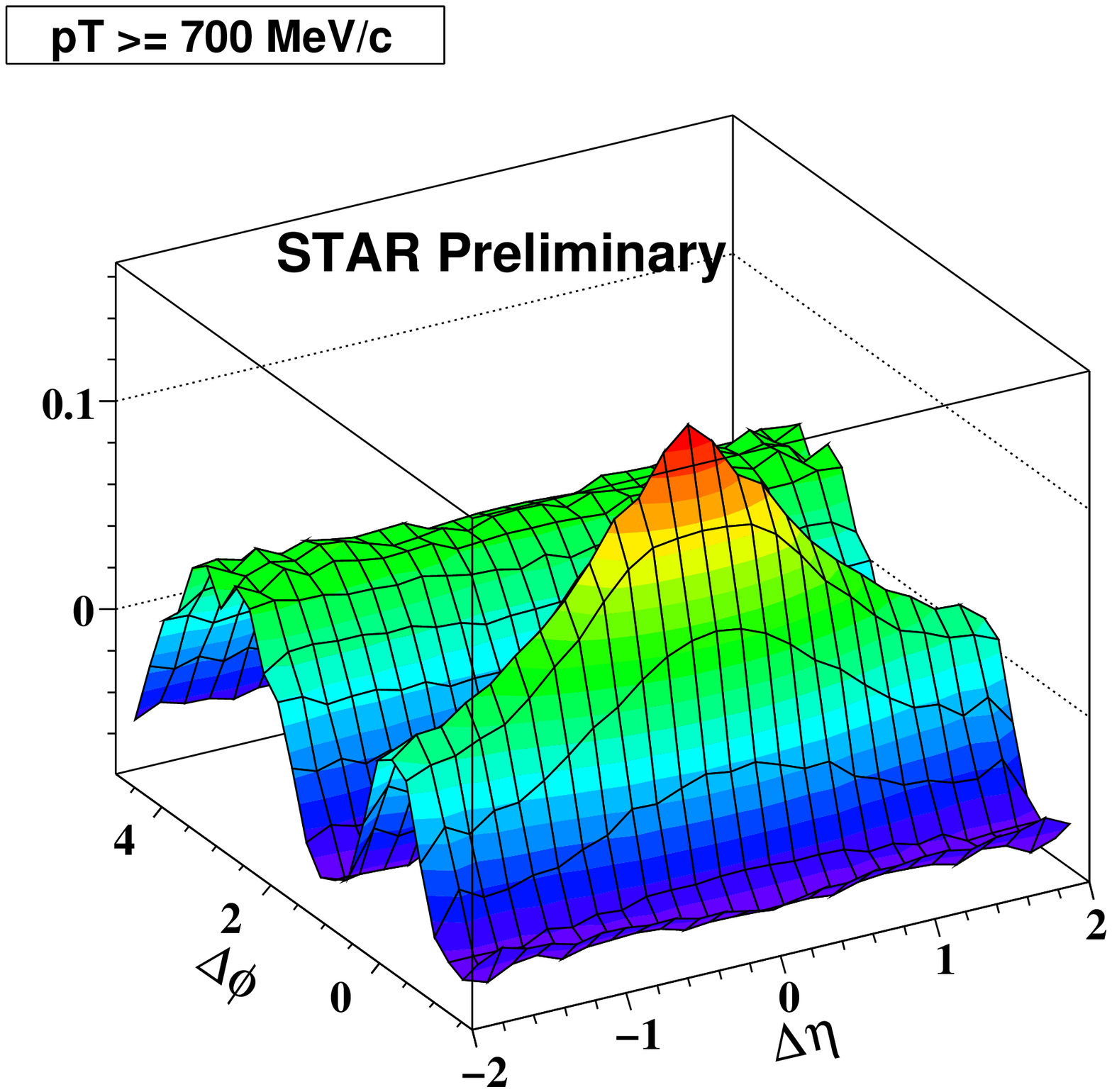}
  \includegraphics{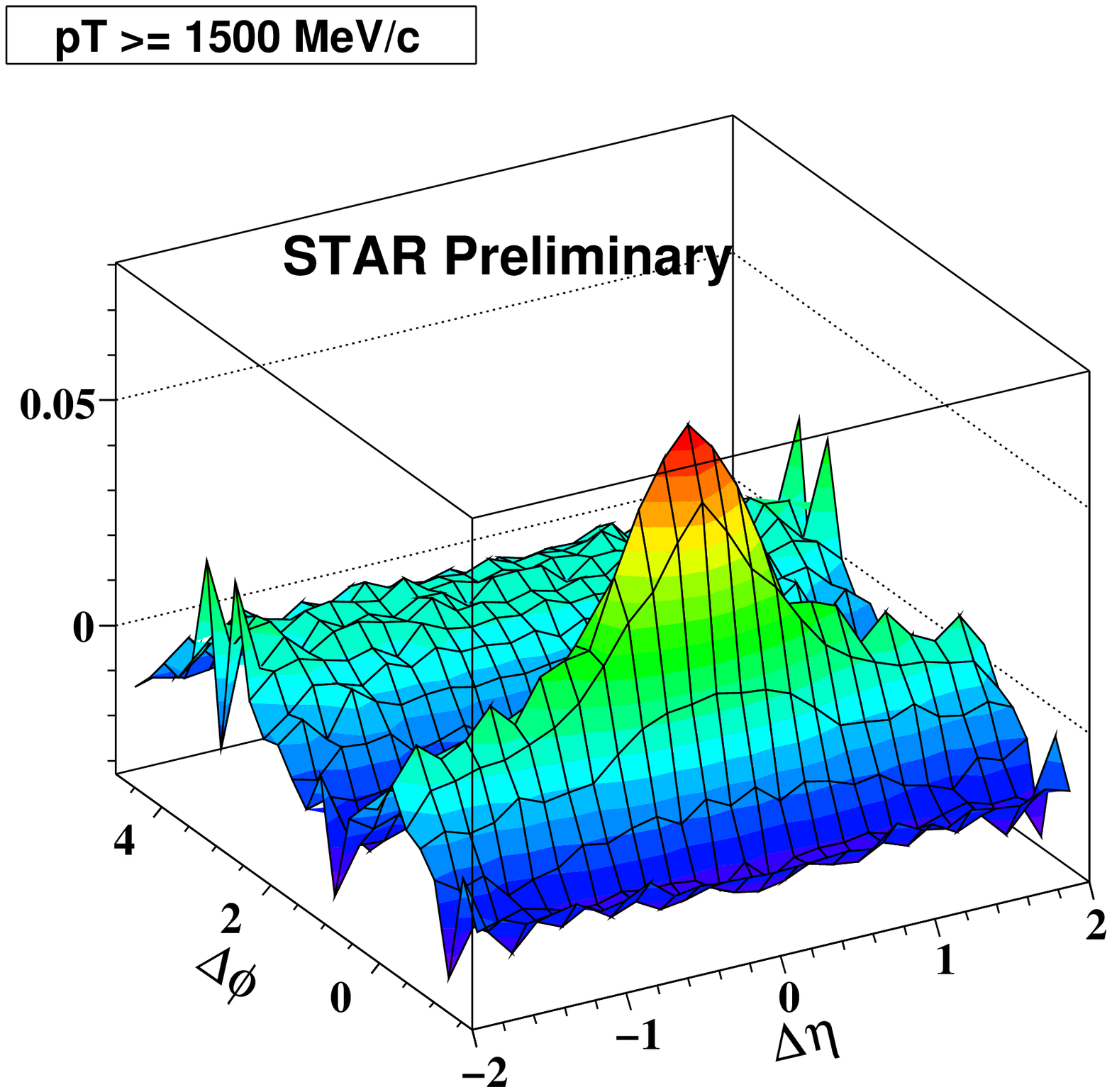}
  }\\

  \caption{Data from \Cu collisions at \sNNtwohundred \cite{DeSilva:2009yy} with \pT cuts gradually applied to an untriggered analysis.  In this analysis both the trigger and the associated particles have the same kinematic cuts applied.  Colour online.}\label{Chanaka}
\end{figure}

Untriggered di-hadron correlations are generally analysed by fitting the signal, including the background term in \Eref{eq:bkgd}.  There is an added term at small (\dphi,\deta) from conversion electrons and HBT correlations which is fit to a 2D Gaussian and in most analyses the \as is fit as a cos(\dphi) term.  The \ridge is parameterized as a 2D Gaussian in untriggered analyses, generally called the ``Soft Ridge''.  While the \ridge observed in \highpT triggered correlations, sometimes called the ``Hard Ridge,`` is independent of \deta within errors, the Soft Ridge clearly has a dependence on \deta \cite{STARMiniJets2006}.  \Fref{Chanaka} shows that the Soft Ridge evolves into the Hard Ridge when a momentum threshold is introduced \cite{DeSilva:2009yy}.  This method is sensitive to whether or not the functional form used in the analysis is a valid description of the data.  The same shape is assumed for the background as in the ZYAM method, making it as sensitive to the validity of the shape of the background as other analyses.  It is less sensitive to event-by-event fluctuations in \vtwo, non-flow contributions to \vtwo, and any correlation of jets with the reaction plane because the fit determines an average effective \vtwo and uses the information that \vtwo is roughly independent of \deta in the region studied.  The cos(\dphi) term assumed to describe the \as in these studies leads to a dip on the \ns.  Since the Mach Cone structure on the \as is present in some kinematic regions before background subtraction \cite{BrianCole}, the assumption that the \as can be described by a cos(\dphi) cannot be valid for all kinematic cuts.  In addition, the \as is clearly not described by a cos(\dphi) in \dAu, \py, or at \highpT.

\subsection{Results}

\begin{figure}
  \centering\resizebox{7cm}{!}{\includegraphics{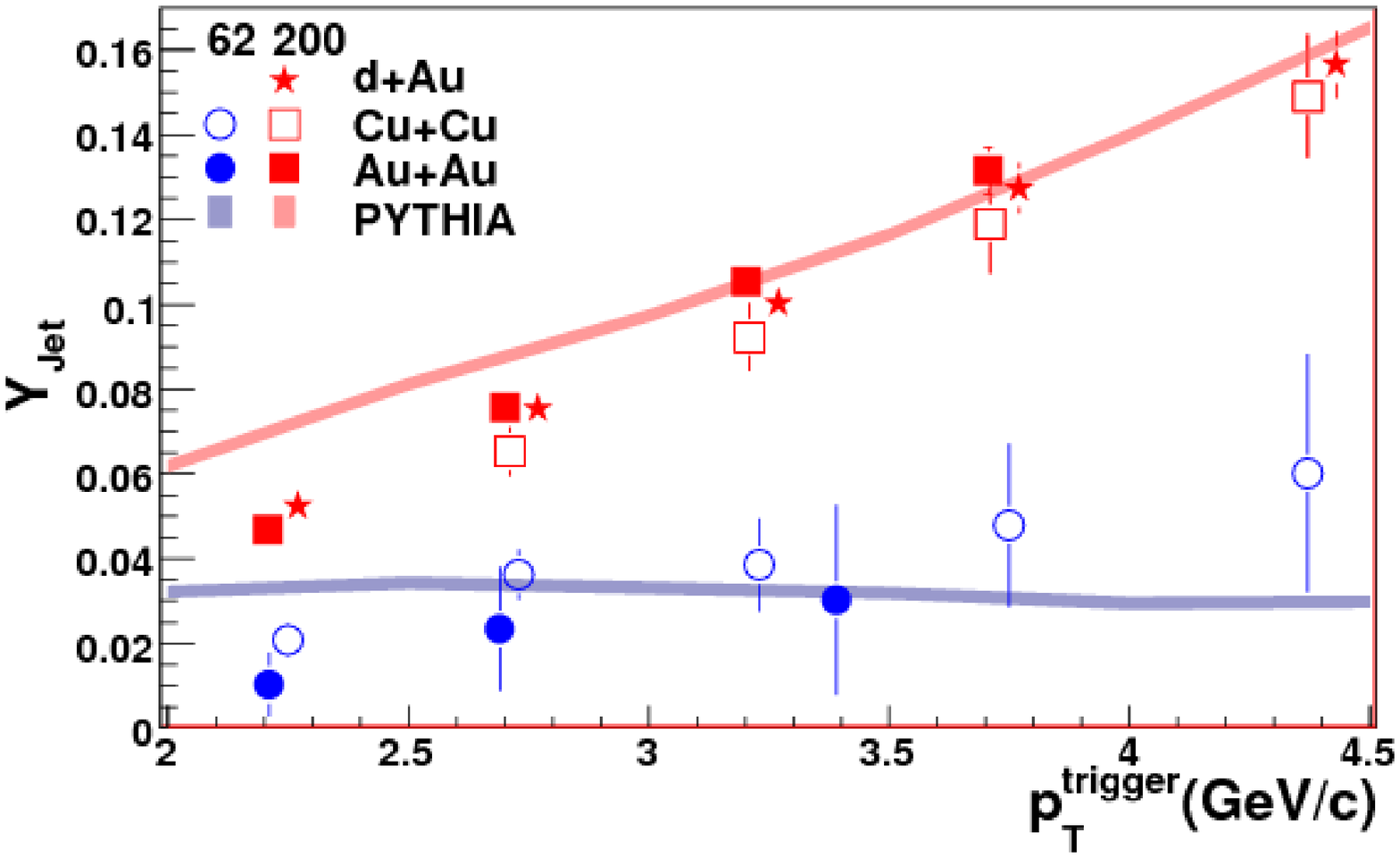}}\\

  \caption{The \pttrig dependence of the \jly per trigger particle for \stdtrig and \stdassoc for minimum bias \dAu, 0-60\% central \Cu, and 40-80\% central \Au collisions at \sNNtwohundred and 0-60\% central \Cu and 0-80\% central \Au collisions at \sNNsixtytwo \cite{Nattrass:2008rs} with comparisons to \py version 6.4.10 \cite{Sjostrand:2006za} tune A \cite{Field:2002vt} at \sNNsixtytwo (blue) and \sNNtwohundred (red) \cite{NattrassUserMeeting}.  Colour online.}\label{TrigPt}
\end{figure}

The \jlc is separated from the \ridge and the background by using the observation that both the \ridge and the \vtwo modulated background are independent of \deta at \highpT, while the \jlc is dependent on both \deta and \dphi.  This means that the \jly is not sensitive to the assumptions made in the ZYAM method.  \Fref{TrigPt} shows the dependence of the \jly, the number of particles associated with a trigger particle, as a function of \pttrig compared to \py version 6.4.10 \cite{Sjostrand:2006za} tune A \cite{Field:2002vt}.  No dependence on the collision system is observed in the data, consistent with the expectation that the \jlc is produced dominantly by fragmentation.  While \py overestimates the yield at lower \pttrig, the agreement is still remarkable given that comparisons are made to \AplusA data.

\begin{figure}
  \centering\resizebox{14cm}{!}{
  \includegraphics{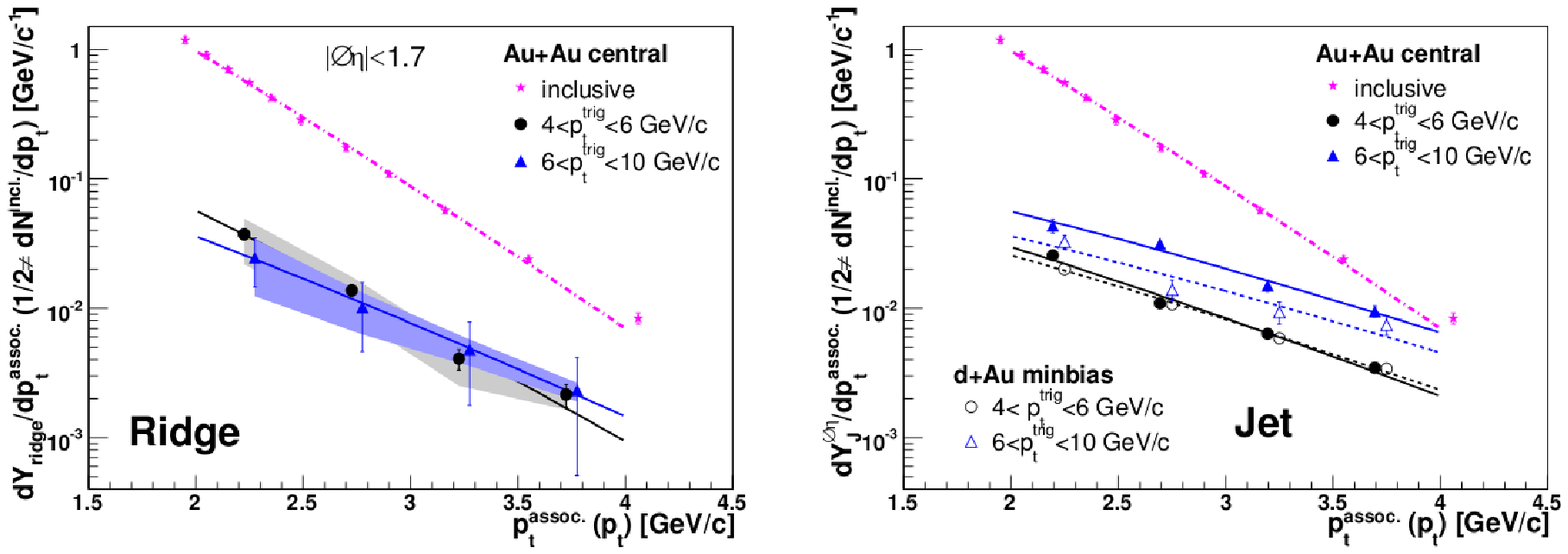}
  }\\

  \caption{Spectra of particles in the \ridge (left) and particles in the \jlc (right) compared to inclusive particle spectra for \assocrangevar{2}.  Figure from \cite{STARRidgePaper}. Colour online.}\label{JoernSpectra}
\end{figure}

The \ridge has been measured at RHIC by STAR \cite{STARRidgePaper,STARMiniJets2006,Agakishiev:2010ur,Abelev:2009jv,Adams:2005ph}, PHOBOS \cite{PHOBOSRidge}, and PHENIX \cite{PhysRevC.78.014901} experiments.  Measurements indicate that the \jlc is dominantly produced by the fragmentation of hard partons, while the \ridge is comparable to the bulk.  The spectra of particles in the \jlc and in the \ridge are shown in \Fref{JoernSpectra}.  The spectra of particles in the \jlc is harder for higher \pttrig, consistent with expectations if the \jlc is dominatly produced by fragmentation.  By comparison the spectra of particles in the \ridge has a slope comparable to the inclusive particle spectra.

\begin{figure}
  \centering
  \resizebox{!}{5cm}{\includegraphics{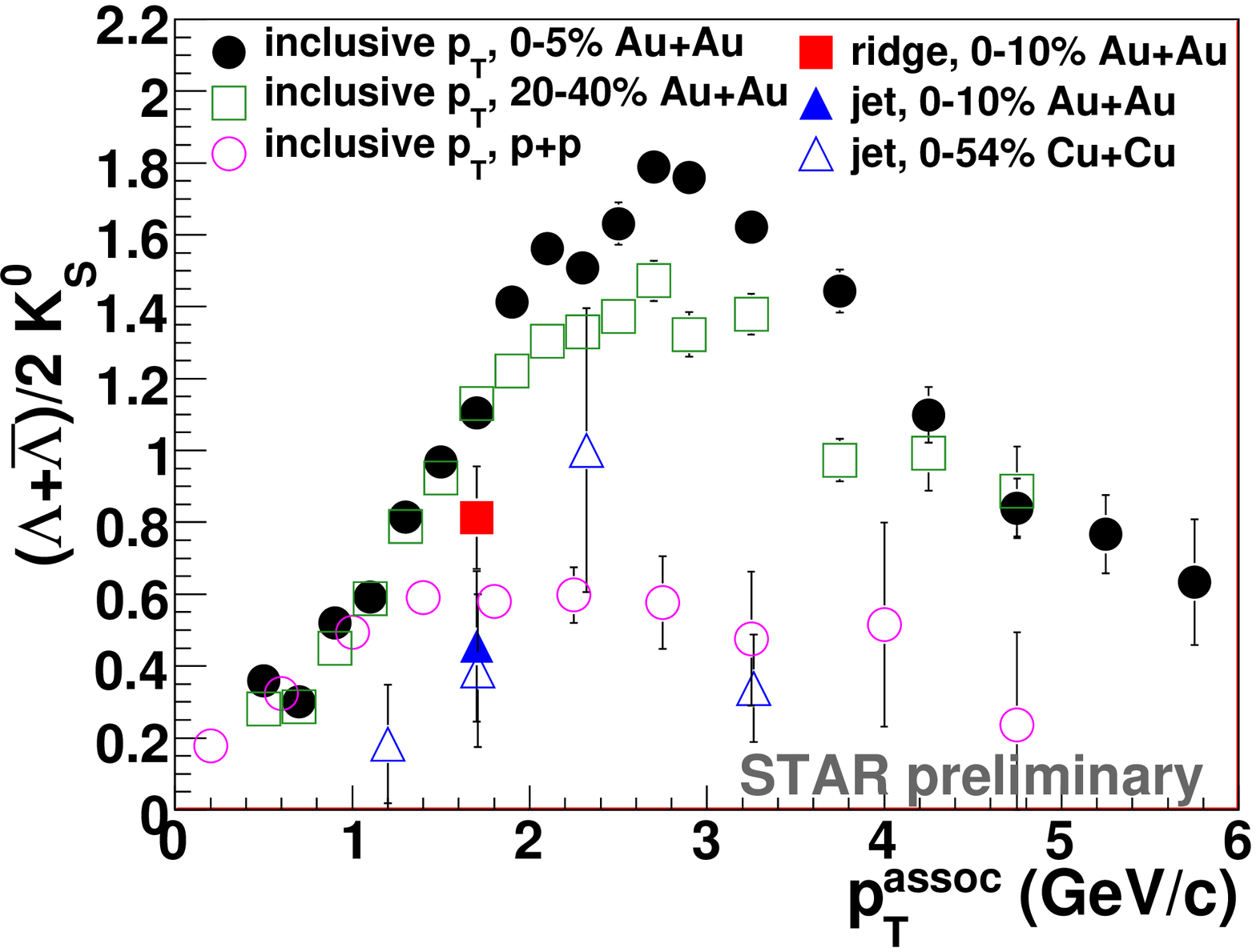}}
  \resizebox{!}{5cm}{\includegraphics{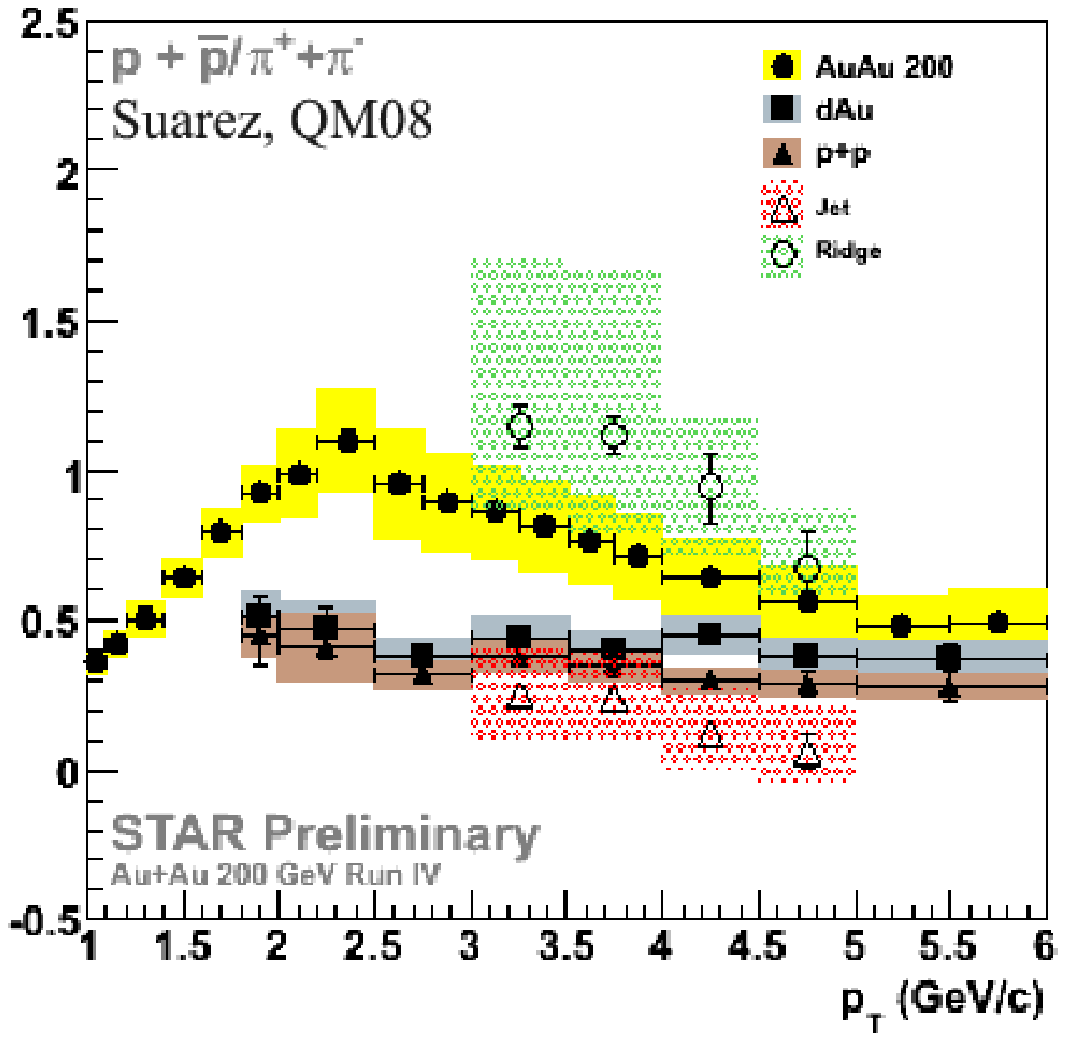}}\\

  \caption{Comparison of particle ratios in the \jlc and in the \ridge for \baryontomeson \cite{Nattrass:2008rs} (left) and for \ptopi \cite{SuarezQM08} (right).  Colour online.}\label{Chemistry}
\end{figure}

Raw correlations clearly show different behavior for both baryons and mesons \cite{PHENIXPartIDCorr2007}.  \Fref{Chemistry} shows the \baryontomeson and \ptopi for the inclusive particle spectra, the \jlc and the \ridge.  The composition of the \jlc is comparable to the inclusive \pp ratios, which are expected to be dominated by jet fragmentation at \highpT.  The composition of the \ridge is comparable to the inclusive spectra for both \baryontomeson and \ptopi.

\begin{figure}
  \centering\resizebox{14cm}{!}{
  \includegraphics{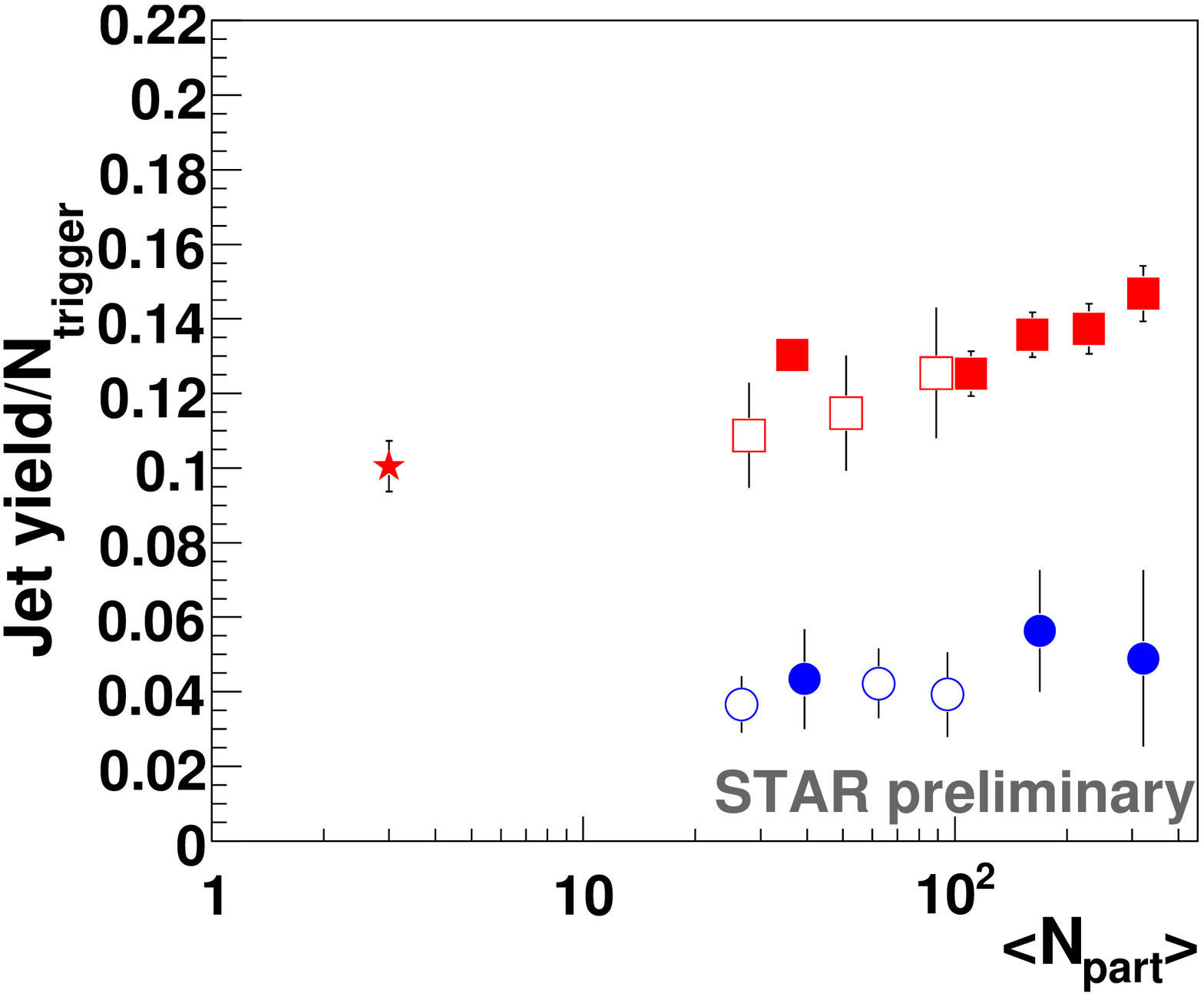}
  \includegraphics{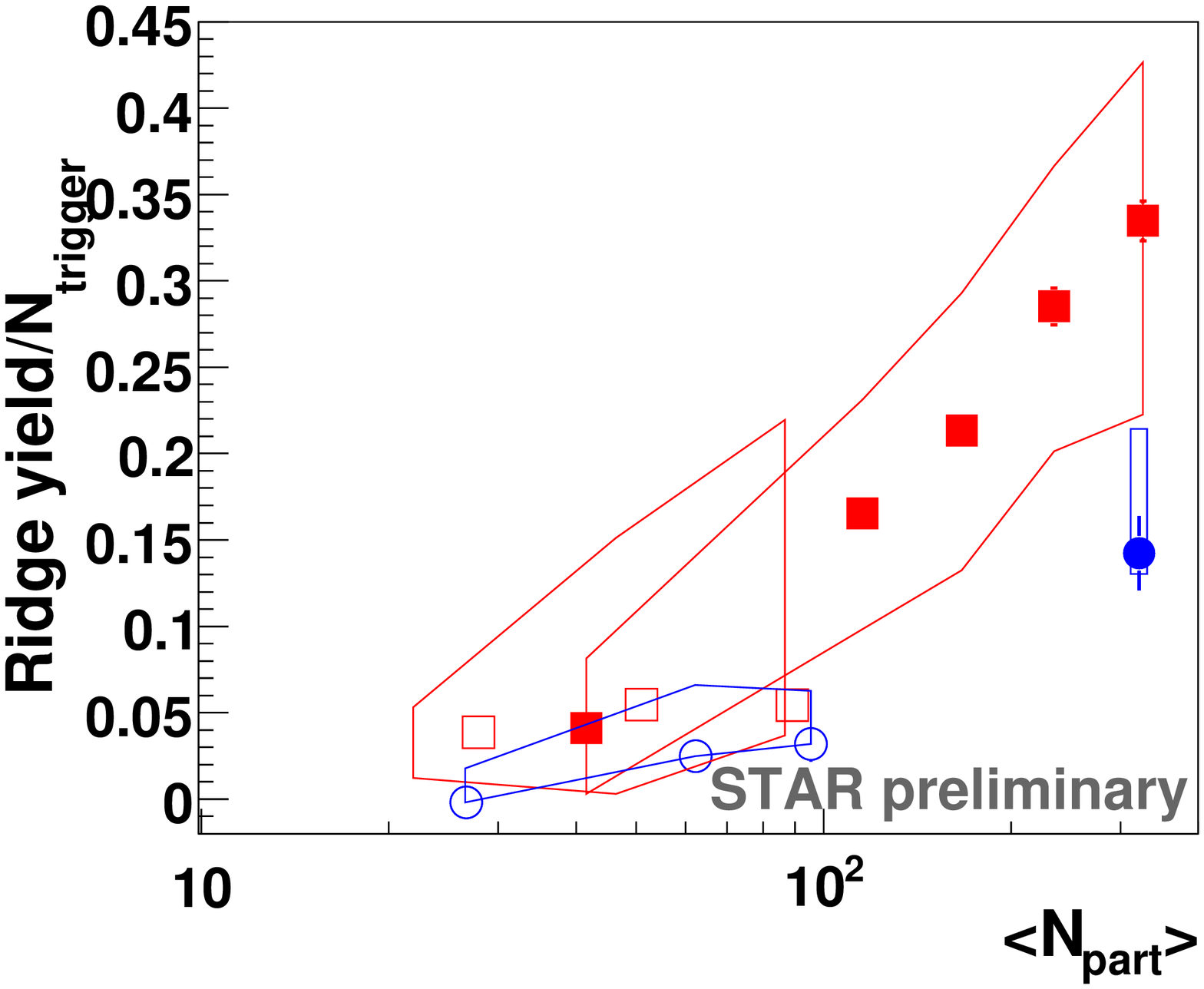}
  \includegraphics{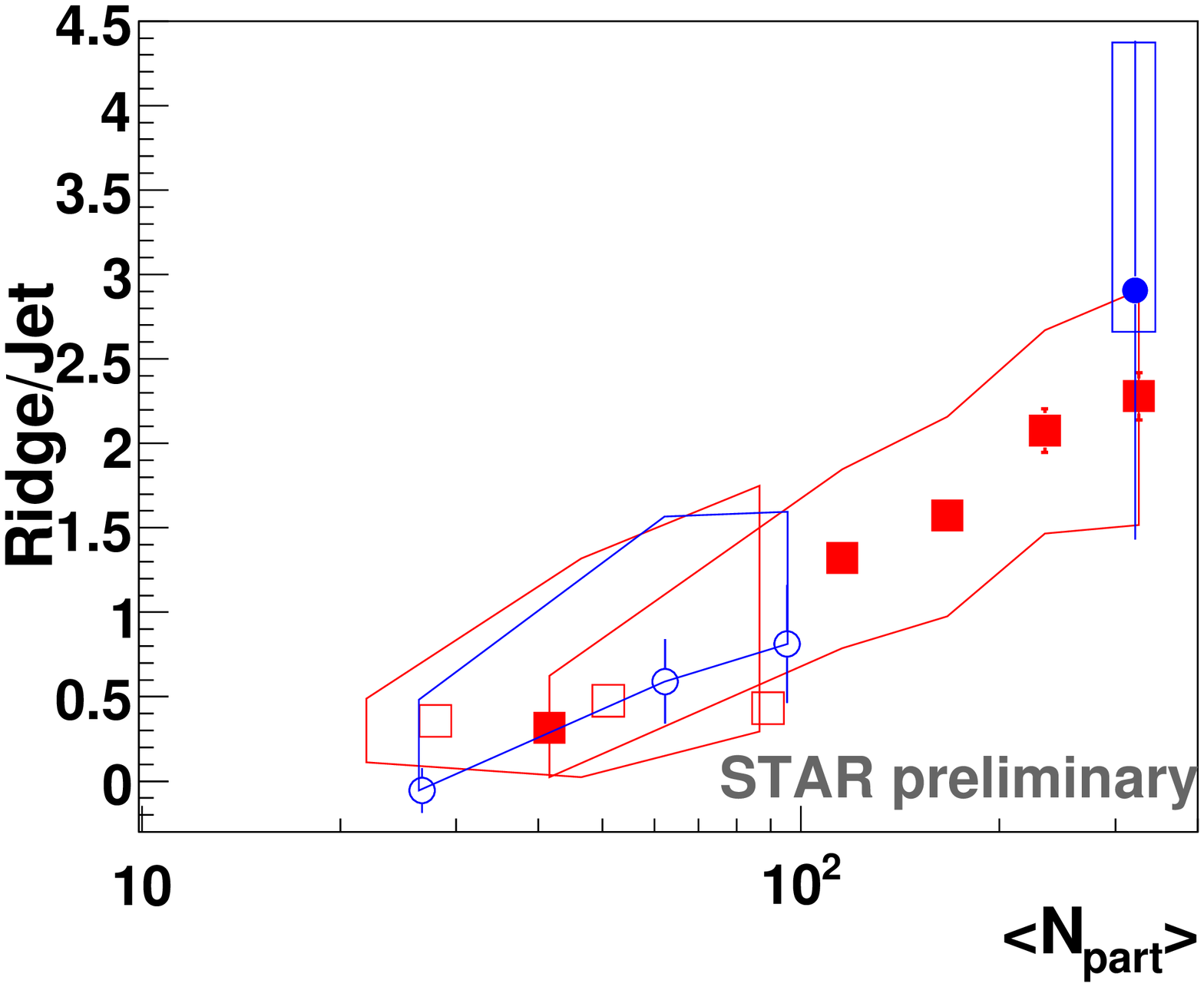}
  }\\

  \caption{Dependence of the \jly (left), \ry (middle), and \jly to \ry ratio (right) per trigger particle for \stdtrig and \stdassoc. Data are from \cite{Nattrass:2008rs}.  Colour online.}\label{Npart}
\end{figure}

\begin{figure}
  \centering\resizebox{7cm}{!}{
  \includegraphics{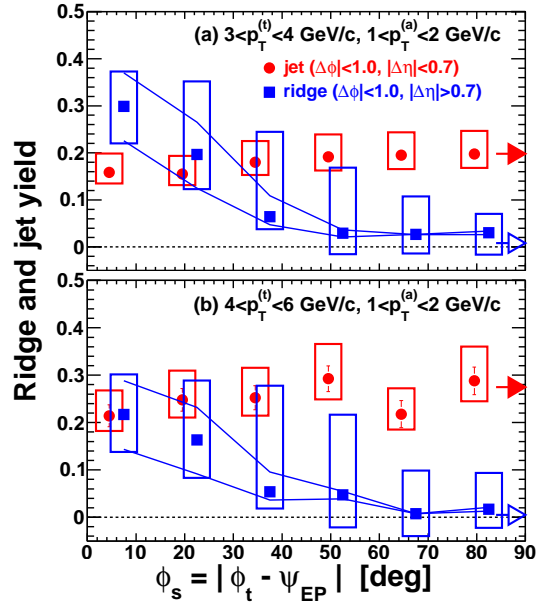}
  }\\
  \caption{The reaction plane dependence of the \jlc and the \ridge.  Data are from \cite{Agakishiev:2010ur}.  Colour online.}\label{Aoqi}
\end{figure}

\Fref{Npart} shows the \jlc, the \ridge, and the ratio of the \jlc to the \ridge as a function of \npart.  The \jly shows litte dependence on the system size over two orders of magnitude in \npart while the \ry increases by a factor of three in the same range.  Both the \jly and the \ry are considerably smaller in \sNNsixtytwo than in \sNNtwohundred, and the ratio of the \jly to the \ry is the same for both energies.  From these data we anticipate the presence of the \ridge in \AplusA collisions both at lower energies in the RHIC beam energy scan and at higher energies in \Pb collisions at the LHC.  \Fref{Aoqi} shows the dependence of the \ry and the \jly on the angle relative to the reaction plane.  The \ridge is clearly dominantly in the reaction plane.


Three particle correlations on the \ns have been studied to determine whether the particles in the \ridge are correlated with each other \cite{Abelev:2009jv}.  These results should be interpreted carefully because of the kinematic limits, however, they indicate that particles in the \ridge are not correlated with each other.  These studies also indicated that while the \jlc is dominantly from charged hadrons with opposite signs, the \ridge is present for charged hadrons with the same sign.

The observations from \AplusA lead to significant experimental constraints on models for the production of the \ridge in \AplusA.  The recent observation of a \ridge in high multiplicity \pp collisions by CMS \cite{PHOBOSRidge} also constrains models, provided the structure observed by CMS and in \AplusA at RHIC arise from the same mechanism.

\section{Models for the ridge}
There are multiple models for the production of the \ridge.  It is useful to break these models down into different classes:
\begin{itemize}
 \item Causal models - the \ridge is created by the interaction of a hard parton with the medium.
 \item Hydrodynamical models - the \ridge is actually a background from hydrodynamical flow.
 \item Initial conditions - the \ridge arises from initial conditions in the incoming nuclei.
\end{itemize}

Causal models include the momentum kick model \cite{Wong:2008yh}, gluon brehmsstrahlung \cite{Armesto:2004pt} and recombination \cite{Chiu:2005ad}.  In the momentum kick model, the \ridge is formed by collisional energy loss of the hard parton with particles in the medium \cite{Wong:2008yh}.  The momentum kick model is consistent with the data, but predicts a sharp drop in the amplitude of the \ridge just outside of the acceptance of the range of the measurements available so far \cite{Wong:2008yh}.  It has also been proposed that the \ridge is formed by gluon brehmsstrahlung in the medium dispersed by flow \cite{Armesto:2004pt}, or through medium heating in combination with recombination of quarks and gluons on the medium \cite{Chiu:2005ad}.  In the gluon brehmsstrahlung model the \ridge would come from fragmentation and therefore would have a similar composition to the \jlc, but this is not observed in the data.  In both the momentum kick model and the recombination model, the \ridge arises from medium partons, leading to a composition similar to the bulk.  However, generally causal models have difficulty producing a \ridge large enough in \deta to be consistent with the data and these models may not be consistent with the data from 3-particle correlations \cite{Abelev:2009jv}.  These models would also imply the existence of a medium in \pp collisions and therefore would be a rather speculative explanation for the CMS data.

There are two main mechanisms for the production of the \ridge which involve hydrodynamical flow.  In the radial flow plus trigger bias model, the \ridge arises because both radial flow and jet quenching lead to the emission of particles from the surface of the medium.  Since both particles from the fragmenting hard parton and from the medium are emitted from the surface of the medium, these particles are correlated in space \cite{Pruneau:2007ua}.  In the $v_3$ model, fluctations in the initial overlap region lead to a non-zero $v_3$ on average.  This $v_3$ leads to a $\cos(3\Delta\phi)$ term in the correlation \cite{Sorensen:2010zq,Alver:2010gr}.  In these models, the \ridge is basically a hydrodynamical background that was not considered.  This easily explains why the \ridge composition is similar to the bulk.  These models are consistent with the reaction plane dependence of the \ridge.  Since there is already considerable evidence for hydrodynamical flow in \AplusA collisions \cite{Voloshin:2008dg}, these effects would be expected and would be a straight forward explanation for the \ridge in \AplusA collisions.  However, these models would also require the existence of a medium in \pp collisions to explain the CMS data.

There are a few different models which explain the \ridge through various initial state conditions.  In a heavy ion collision there are large QCD magnetic fields early in the collision and the large fluctuations in these QCD magnetic fields can lead to a \ridge \cite{Romatschke:2006bb}.  This production mechanism may also be present in \pp collisions and may be able to explain the \ridge in \pp without any need for a medium or hydrodynamical flow.  However, it is not clear that this mechanism can produce a \ridge large enough to explain the \AplusA data.  In addition, it has been proposed that hot spots early in the collision could lead to fluctuations which could explain the \ridge \cite{CarstenGreiner}.  It is not clear what could lead to these hot spots in \pp collisions that may be able to explain the CMS data.

It is not straightforward to distinguish between various production mechanisms for the \ridge because some calculations include a combination of initial conditions which may lead to fluctuations and hydrodynamical effects.  In a full hydrodynamical calculation, both the radial flow plus trigger bias mechanism and $v_3$ may lead to a \ridge.  It is possible for the \ridge in \pp and in \AplusA to arise from different mechanisms, but it would be simpler if the \ridge were produced by the same mechanism in both \pp and \AplusA collisions.  A better theoretical understanding of the models for the production mechanism for the \ridge is needed in order to understand the effects each model would predict.


\section{Conclusions}
There are extensive data on the \ns of di-hadron correlations from \AplusA collisions.  The data indicate that the \jlc arises dominantly from fragmentation of a hard parton, perhaps with some modification.  The \ridge is well characterized in \AplusA and sufficient data are available to constrain models.  Several additional experimental constraints will be available in the near future.  The beam energy scan at RHIC allows studies of the \ridge at lower energies \cite{Aggarwal:2010cw} and the recent \Pb collisions at the LHC will enable studies of the \ridge at higher energies.  In addition, studies of the \ridge in \pp at the LHC could be useful for constraining models.  The simplest explanation would be one that could explain both the \pp and the \AplusA data.  Studies of the charge dependence and particle composition of the \ridge in \pp collisions may help clarify whether the \pp \ridge and the \AplusA \ridge are produced by the same mechanism.  Searches for the \ridge in high multiplicity collisions at lower energies, particularly those where the \ridge was observed in \AplusA collisions, would be interesting.  However, more quantitative comparisons with existing data could also considerably constrain models.


\bibliographystyle{varenna}
 \bibliography{Bibliography}

\end{document}